\documentclass[article,twocolumn,prl,showpacs]{revtex4-1}  
\usepackage{graphicx,amsmath,mathrsfs}
\usepackage{amssymb}
\usepackage{amsthm,multirow}
\usepackage{bm,bbm}
\usepackage{subfigure,xcolor} 
\usepackage{color}
\mathchardef\Gamma="7100

 \def\bq{{\bf q}}

\def\bx{{\bf x}}

\def\bp{{\bf p}}

\def\bz{{\bf z}}
\def\bj{{\bf j}}
\def\bd{{\bf d}}

\def\bA{{\bf A}}

\def\cL{{\cal L}}

\def\cP{{\cal P}}

\def\cO{{\cal O}}
\def\cW{{\cal W}}
\def\cM{{\cal M}}

\def\Im{{\mbox{Im}}}
\def\Re{{\mbox{Re}}}
\def\Tr{{\mbox{Tr}}}

\def\be{\begin{equation}}
\def\ee{\end{equation}}
\def\bea{\begin{eqnarray}}
\def\eea{\end{eqnarray}}

\begin{document}
%--------------------------------------------------------------------------------
\title{The Hall Number of Strongly Correlated Metals}

\author{Assa Auerbach}
\affiliation{Physics Department, Technion, 32000 Haifa, Israel}
 
\date{\today }

\begin{abstract}
An exact formula for the temperature dependent Hall number of metals is derived. It is valid for non-relativistic fermions or bosons, with arbitrary potential and interaction.  
This DC transport coefficient is proven to (remarkably) depend solely on equilibrium susceptibilities, which are more amenable to numerical algorithms than the conductivity. 
An application to strongly correlated phases is demonstrated by calculating the Hall sign in the vicinity of Mott phases of lattice bosons.

%For weak disorder, high order terms yield corrections to Drude's result.  
%For strong potentials and interactions, effective Hamiltonians can be used.  Hall sign changes of lattice bosons near the Mott phases are obtained. 
%Future applications are proposed.
 \end{abstract}

\maketitle
 The zero field Hall number or ``carrier density'' of a metal  is defined by
\be 
n_H \equiv - \left(  {d\rho_{H}\over dB}\Big|_{B=0}  e^* c\right)^{-1}.
\label{nH}
\ee
$\rho_{H},B,e^*,c$  are the (magnetic field antisymmetric) Hall resistivity, magnetic field, quasiparticle charge and speed of light respectively. 
This definition is rooted in Drude-Boltzmann\cite{ziman}  theory for weakly interacting conduction electrons (holes) of density $n$, and charge $e^*\!=\!e$ ($-e$).
While  {\em conductivities} $\sigma_{xx},\sigma_{H}$, depend on the quasiparticles' effective mass and scattering time,   
for isotropic Fermi liquids,  these properties cancel out in ${d\rho_{H}\over dB}\!=\! -\sigma_{xx}^{-2} {d\sigma_{H}\over dB}\!=\!-1/(n e^*c)$.  

The experimental Hall number, however, has defied a ``carrier density''  interpretation in strongly correlated metals. In the normal phase of cuprates~\cite{hagen,Taillefer} and in disordered superconducting films~\cite{smith,kapitulnik} $n_H$ exhibits anomalous  temperature dependences, 
and sign changes, which 
have posed a challenge to theory\cite{KKS}. When quasiparticles' scattering rate is too high,
Boltzmann transport theory has questionable validity.

For gapped phases and finite lattices $\sigma_{xx}\!=\!0$, and  $\rho_{H}\!=\!-\sigma_{H}^{-1}$ can be calculated by Chern numbers on the torus\cite{yosi,lindner,huber,berg}, however, computing  both $\sigma_{xx},
\sigma_{H}$ in the resistive phases ($\sigma_{xx}\!>\!0$) of strongly correlated systems, {\em is
notoriously difficult:}
Diagrammatic expansions of the Kubo formulae require infinite resummations\cite{mahan}. 
Exact diagonalization suffers from small lattice sizes~\cite{Lanczos,ED},  quantum Monte Carlo simulations~\cite{prokofiev} from ill-posed analytical 
continuation~\cite{MaxEnt,Snir}, and continued fraction calculations~\cite{viswanath,lindner,khait}
require extrapolation schemes. Approximations for $d\rho_{H}/dB$ include high frequency~\cite{SSS}, retraceable paths~\cite{brinkman}, Drude weight derivatives~\cite{Prelovsek} , and dynamical mean field
theory~\cite{Lange}.  However an {\em exact} (generally valid), computable expression is in dire need.

In this paper, I derive  a summation formula, given by Eq.~(\ref{Formula}), for the temperature dependent Hall number of
non-relativistic fermions or bosons, in an arbitrary potential and two-body interaction strength. 
Remarkably, the formula expresses a DC transport coefficient solely in terms of equilibrium susceptibilities. Such a property of the Hall number was previously suggested,
but not proven, except in the high frequency limit~\cite{SSS}. Susceptibilitites are much more amenable to numerical computation than the conductivity, which miraculously drops out of the Hall number.  
Properties of the magnetic Liouvillian in Bogoliubov hyperspace are essential in the derivation.
The leading term in the sum recovers Drude-Boltzmann's result at weak disorder.
For strong lattice potentials and interactions, projected Hamiltonians may be used to compute the susceptibilities.
As an example, I evaluate the Hall sign for strongly interacting lattice bosons.  The results extend previous Chern number calculations\cite{lindner,huber} to finite temperatures. Future applications are discussed.

{\em Hamiltonian and Kubo Formulae}. --- We consider $N$ interacting particles in volume $V$ in an arbitrary bounded potential $\Phi$,
\be
H = \sum_{i=1}^N {\left(\bp_i-{e^*  \over c} \bA(\bx_i) \right)^2\over 2m}   + \Phi(\bx_i) + {1\over 2}\sum_{i\ne j} U(|\bx_i-\bx_j|) .
\label{Schrod}
\ee
$\bA(\bx)=  {B\over 2}(\hat{\bz} \times \bx )$. The zero wave vector current operators
are  $j^\alpha  ={e^* \over m} \sum_i (p^\alpha_i-{e^*  \over c}A^\alpha(\bx_i))$. 

The Bogoliubov hyperspace of operators is defined by inner products\cite{Bogoliubov,forster,MerminWagner}. For any two operators (hyperstates) $A,B$,
 \be
\left(A|B\right) =  {1\over Z} \sum_{n\ne m} { e^{-\beta E_n} - e^{-\beta E_m}\over E_m-E_n}  A^*_{mn}B_{mn}
\label{product}
\ee
where $E_n$ is the spectrum of $H$, and $Z$ is the partition function. $(A|B)$ is a thermodynamic susceptibility. 
In this hyperspace, the Liouvillian  $\cL\equiv [H,\bullet]$ is a hermitian hyperoperator,  and $\bullet$ is any operator. The Liouvillian resolvent $\left({1\over \cL-i0^+} \right) \!\equiv\! \left({1\over \cL} \right)' + i \left({1\over \cL} \right)'' $, separates into the hermitian and antihermitian parts. (The latter's eigenvalues are energy conserving delta functions.)
The DC conductivities\cite{Comm-limits} are written in hyperspace notation as (for the derivation see Supplementary Material\cite{SM} (SM)), 
\bea
\sigma_{xx}  &=&{\hbar\over V} \Re \left( j^x  \Big|\left( {1\over \cL }\right)'' \Big| j^x\right),\nonumber\\
\sigma_{H}  &=&{\hbar\over V} \Im  \left( j^x  \Big| \left( {1\over \cL}\right)' \Big| j^y\right).
\label{cond}
\eea

Defining $\rho = e^{-\beta H}/Z$  the operators can be reorganized as\cite{Com-Yosi},
\be
\sigma_{H} =- {\hbar\over V} \Im \Tr  \left\{ \rho \left[  \left({1\over \cL}\right)' j^x ,  \left( {1\over \cL}\right)'  j^y\right]\right\}.
\ee
Differentiating the density operator yields,
\be
{d\rho\over dB} = -\left[ \rho, \left({1\over \cL}\right)' M \right]- \beta \langle M\rangle ,
\label{drho}
\ee
and differentiating the resolvent yields,
\be
{d\over dB} \left( {1\over \cL}\right)' =  \left( {1\over \cL }\right)'  \cM    \left({1\over \cL} \right)'  - \left( {1\over \cL }\right)''  \cM    \left({1\over \cL} \right)'', 
\label{dcL}
\ee
where
\be 
M \equiv -{ dH\over dB},\quad \cM\equiv [ M,\bullet],
\label{cM}
\ee
are the magnetization operator, and magnetization hyper-operator, respectively.

The field derivative of the Hall conductivity\cite{Com-B} is given by a sum of five terms:
 \be
{ d\sigma_{H} \over dB} \Big|_{B=0} =      \Xi_{\rm osc} + \Xi_{\rm comm} +   \Xi_{j} + \Xi_\cM' + \Xi_\cM''  . 
\label{terms}
\ee
I shall now show that the sum over the first 4 terms in (\ref{terms})  vanishes identically.

The first term, using (\ref{drho}), is
\bea
&& \Xi_{\rm osc}=  {\hbar\beta\over V}\Im \Tr \left\{ \rho  M^{\rm diag}   \left[ \left({1\over \cL}\right)'  j^x, \left({1\over \cL}\right)'   j^y \right] \right\} \nonumber\\
 && ~~~~~~~ ~~~~~~~~~~~~~~~~~~~~~~~~~~~~- \beta \langle   M^{\rm diag}\rangle~ \sigma_{H}=0 ~,
\eea
where $M^{\rm diag}$ is the energy-diagonal part of $M$, which vanishes at zero field.

The other terms, using (\ref{drho},\ref{dcL}), are,
 \bea
 && \Xi_{\rm comm} =  {\hbar\over V} \Im \left( M \Big|   \left[ \left({1\over \cL}\right)'  j^x, \left({1\over \cL}\right)'   j^y\right] \right), \\
 &&\Xi_j =  {\hbar\over V} \Im\left( \left(  {dj^x\over dB} \Big| \left({1\over \cL}\right)' \Big| j^y  \right) +    \left(  j^x \Big| \left({1\over \cL}\right)'  \Big|{dj^y\over dB}    \right)\right), \nonumber\\ \\
&& \Xi_\cM' =  {\hbar\over V} \Im \left(  j^x \Big|  \left( {1\over \cL }\right)'  \cM    \left({1\over \cL} \right)' \Big| j^y  \right) -  \left( j^x \leftrightarrow j^y\right) , \nonumber\\ \\
 &&\Xi_\cM'' = - {\hbar\over V} \Im \left(  j^x \Big|  \left( {1\over \cL }\right)''  \cM    \left({1\over \cL} \right)'' \Big| j^y  \right) +  \left( j^x \leftrightarrow j^y\right) . \nonumber\\ 
\label{dsigmaXY}
\eea

The following identities hold for $H(B\!=\!0)$:
\bea
&&\left({1\over \cL}\right)'    \bj =  { ie^* \over  \hbar } \bd,\nonumber\\
&&\left[ M,\bd \right] = - i { \hbar e^*  \over 2  mc} \hat{\bz}\times \bd\nonumber\\
&&{ d\bj \over dB} = - {(e^*)^2\over 2mc}   \hat{\bz}\times \bd,
\label{identities}
\eea
where $\bd=\sum_i \bx_i$ is the total polarization operator.

$\Xi_{\rm comm} =0$ since the two polarizations commute,
\be 
\left[ \left({1\over \cL}\right)'  j^x, \left({1\over \cL}\right)'   j^y\right]  =-\left( {e^*\over \hbar}\right)^2  \left[\bd^x, \bd^y \right] = 0 .
\label{scomm}
\ee

It also follows from  (\ref{identities}), that the next two terms cancel each other,
 \bea
\Xi_j &=&   {  (e^*)^3 \over 2mc  V }\Re  \Big( \left( d^x| d^x\right) +   \left( d^y| d^y \right)\Big)    \nonumber\\
 \Xi_\cM'   &=&   {(e^*)^2 \over \hbar V} \Im  \left( d^x| [M,d^y]\right)  -   (x\leftrightarrow y) = - \Xi_j .\nonumber\\
\label{sj+smp}
\eea
Thus we are left with just $\Xi_\cM''$,   
\be
 {d\sigma_{H}\over dB}\Big|_{B=0}  =- {\hbar\over V} \Im \left(  j^x \Big|  \left( {1\over \cL }\right)''  \cM    \left({1\over \cL} \right)'' \Big| j^y  \right) +  \left( j^x \leftrightarrow j^y\right) .  
 \label{dsdB}
 \ee

\begin{figure}[!b]
\begin{center}
\includegraphics[width=7.5cm,angle=0]{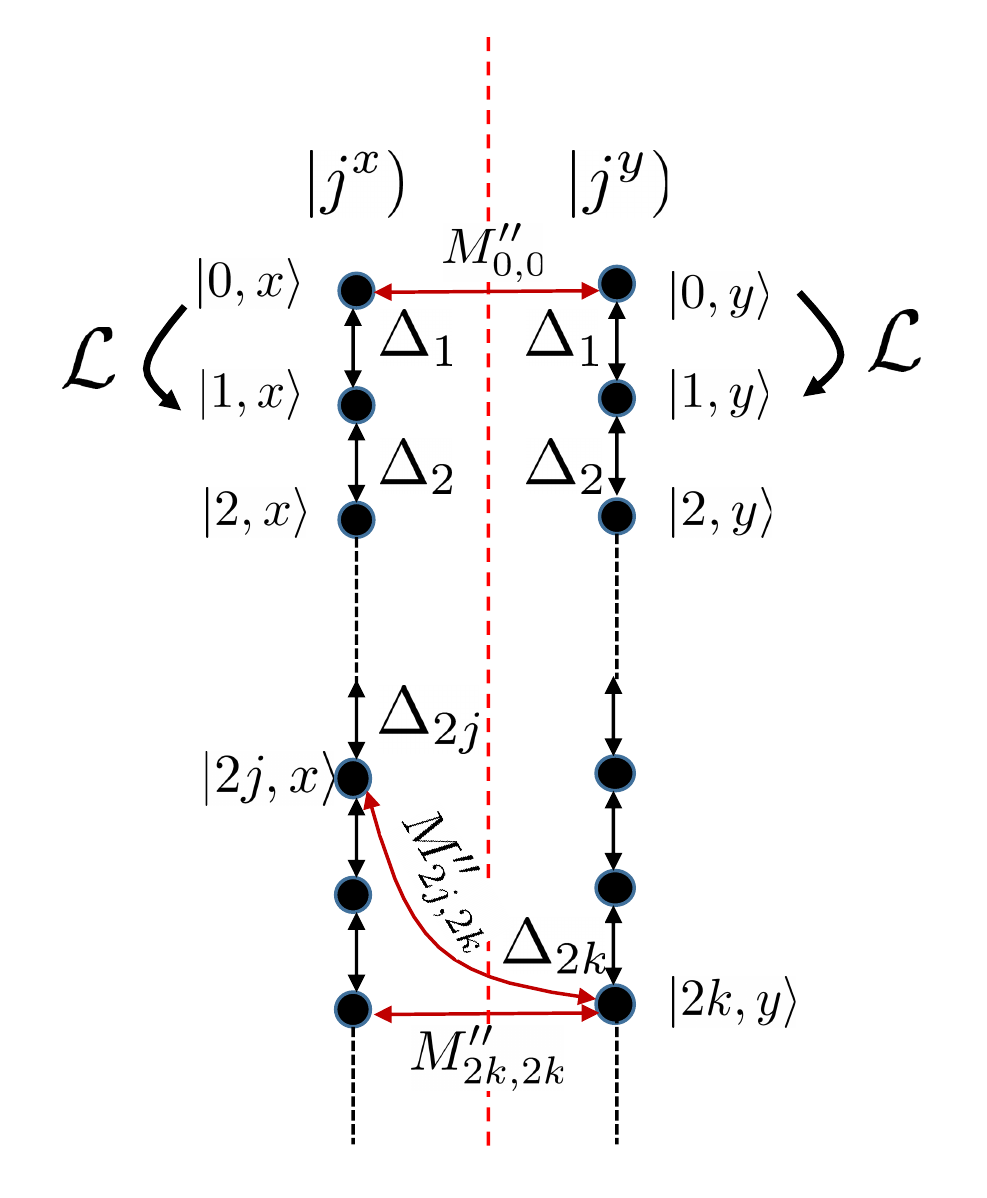}
\caption{The orthonormal Krylov bases, Eq.~(\ref{Krylov}), constructed (for $B\!=\!0$) from $j^x$ and $j^y$ by repeated application of the Liouvillian $\cL$. 
$\Delta_n$ are the recurrents of $\sigma_{xx}$. 
$M''_{n,m}$ are the magnetization matrix elements defined  in Eq.~(\ref{sxy-sum}).
}
\label{fig:Krylov}
\end{center}
\end{figure}

{\em Krylov states and recurrents} --- We set $B\!=\!0$. $H$ is assumed to have  $x\!\leftrightarrow\!y$ symmetry, for simplicity.  Two orthonormal Krylov bases
$|n,\alpha\rangle$,  $\alpha=x,y$ are  constructed,
\bea
&& |0,\alpha\rangle  \equiv { |j^\alpha )\over ( j^\alpha|j^\alpha)^{1\over 2}}~,\nonumber\\
&& |n,\alpha)   \equiv  (1-\cP_{n-2,\alpha}) \cL (1-\cP_{n-3,\alpha})\cdots \cL|0,\alpha\rangle\nonumber\\
&& |n,\alpha\rangle  ={1\over N_n} |n,\alpha),
\label{Krylov}
\eea
where $|\bullet)$  ($|\bullet\rangle$) denote unnormalized (normalized) hyperstates, where $N_n$ are the normalizations of $|n,\alpha)$.   $\cP_{n\alpha}\!=\!|n,\alpha\rangle\langle n,\alpha|$ are projectors.

In Krylov space, the Liouvillian acts as a  hopping Hamiltonian on two semi-infinite chains, as shown in Fig.~\ref{fig:Krylov}, 
\be
L_{n',n}=\delta_{n',n+1} \Delta_n +\delta_{n',n-1} \Delta_{n-1}.
\label{Lnn}
\ee
$\Delta_n = \langle n\!+\!1,\alpha| \cL|n,\alpha\rangle$ are the {\em recurrents}~\cite{viswanath}. The
conductivity moments  $\mu_{2k} \! = \! \hbar^{-2k} V^{-1} (\cL^k j^x | \cL^k  j^x)$,  are computable as thermodynamic susceptibilitites.
$\Delta_n$ is obtained directly 
from $\mu_{2k}$  by the recursive relations~\cite{SM} $\mu_{2k} \!=\!  \hbar^{-2k}  \tau_{xx} \left(L^{2k}[\Delta]\right)_{0,0}$, which depend only on  $\Delta_n,n=1,2\ldots k$. 

The spectral matrix $G''_{0,0}= \Im (i0^+  -L)^{-1}_{0,0}$ yields the continued fraction representation~\cite{lindner},
 \bea
\sigma_{xx} &=&  -\hbar\tau_{xx} G''_{0,0}\nonumber\\
&=& -\hbar \tau_{xx} \Im  { 1\over i0^+ - {  |\Delta_1|^2\over   i0^+ - {  |\Delta_2|^2\over  i0^+ - {  |\Delta_3|^2\over \ldots}} } } ,
\label{sxxG}
\eea
where,
\be
\tau_{xx} ={1\over V}  \left(  j^x | j^x \right) = \int_{-\infty}^\infty\! { d\omega\over \pi} \sigma_{xx}(\omega),
\label{tauxx}
\ee
is the ``$f$-sum rule''.
While computation of low order recurrents is commonly feasible, determination of $\sigma_{xx}$ requires extrapolation~\cite{viswanath,khait} of  $\Delta_n$ to $n\to \infty$, 
a procedure which can suffer from some ambiguity. I  will now show that fortunately, $\sigma_{xx}$  drops out of the Hall number.

{\em Summation formula for $n_H$} -- Inserting (partial) resolutions of identity $1\!=\!  \sum_n \cP_{n,\alpha}$  between the hyper-operators in $\Xi_\cM''$ of (\ref{dsigmaXY}) leads to the following sums,
\bea
{d\sigma_{H}\over dB}\Big|_{B=0}  &=& - {\hbar \tau_{xx} \over V}  \sum_{n,m} G''_{0,n}G''_{m,0} M''_{n,m},\nonumber\\
M''_{n,m}&\equiv& \Im \left( \langle n,x| \cM|m,y\rangle  -  \langle n,y| \cM|m,x\rangle\right) .
\label{sxy-sum}
\eea
All the odd terms $G_{0,2j+1}$ are purely real\cite{SM}, and do not contribute to $\Xi_\cM''$, while the even terms are given by,
\bea
G''_{0,2j} &=&  G''_{0,0} R_j = -{\sigma_{xx}\over \hbar \tau_{xx} }R_j ,\nonumber\\
R_j &\equiv& \prod_{i=1}^j \left( - {\Delta_{2i-1}\over \Delta_{2i} } \right)  .
\label{defs}
\eea
Assuming  a metal with time reversal symmetry, $\sigma_{xx}\!>\!0$, and $d\sigma_{xx}/dB|_{B\!=\!0}\! =\!0$,
one can write 
\be
{d\rho_{H}\over dB}\Big|_{B=0} = -   \sigma_{xx}^{-2}  {d \sigma_{H}\over dB}\Big|_{B=0} 
\ee
Hence, by (\ref{nH}), (\ref{sxy-sum})and (\ref{defs}),  the prefactor of $\sigma_{xx}^{-2}$ is eliminated, and we arrive at,
\bea
&&{1\over n_H}  =  {1\over n_H^{(0)}}+ { e^* c \over  \hbar \tau_{xx}}  \sum_{j,k=1}^\infty   R_jR_k M''_{2j,2k},\nonumber\\
&&{1\over n^{(0)}_H } =  { e^* c \over  V  \hbar \tau^2_{xx}}   \Im \Big( (j^x | \cM|j^y ) -  ( j^y | \cM|j^x ) \Big) .
 \label{Formula}
 \eea

{\em Discussion.} --- Eq.~(\ref{Formula}) is the key result of this paper.  Since for a non critical metal, $|d\rho_{xy}/ dB|<\infty$, this is a conditionally convergent sum.
When truncated, a finite 
subset of recurrents $\Delta_{n}$, and magnetization matrix elements $M''_{n,m}$ need to be computed. The truncation error may be estimated by various perturbative methods, depending on the Hamiltonian, or numerically. {\em Remarkably, all coefficients depend solely on static thermodynamic susceptibilities
as defined by  (\ref{product})}.  Hence they are amenable to well controlled algorithms. A partial list is:\\
(i) Quantum Monte Carlo simulations\cite{prokofiev,assaad} (for sign free models) which compute  imaginary time correlators,
$(A|B) \!=\! \int_0^\beta d\tau  \langle A^\dagger(\tau) B \rangle$.\\
(ii) High temperature series expansion\cite{domb}.\\
(iii) Variational methods, including Density Matrix Renormalization Group\cite{DMRG}, which can compute
$(A|B) = - { \partial^2 F \over \partial h_A \partial h_B}$, where 
 $F$ is a variational free energy which includes the source terms $-h_A A^\dagger - h_B B$.\\
(iv) Eq.~(\ref{product}) may be computed by exact diagonalization on finite clusters, whose linear length
exceeds the correlation length.  We note that exact diagonalizations are problematic when approaching e.g. superconducting, magnetic or charge density wave instabilities.

Formula  (\ref{Formula}) will now be demonstrated  for weak and strong interaction models.

\begin{figure}[!t]
\begin{center}
\includegraphics[width=8.5cm,angle=0]{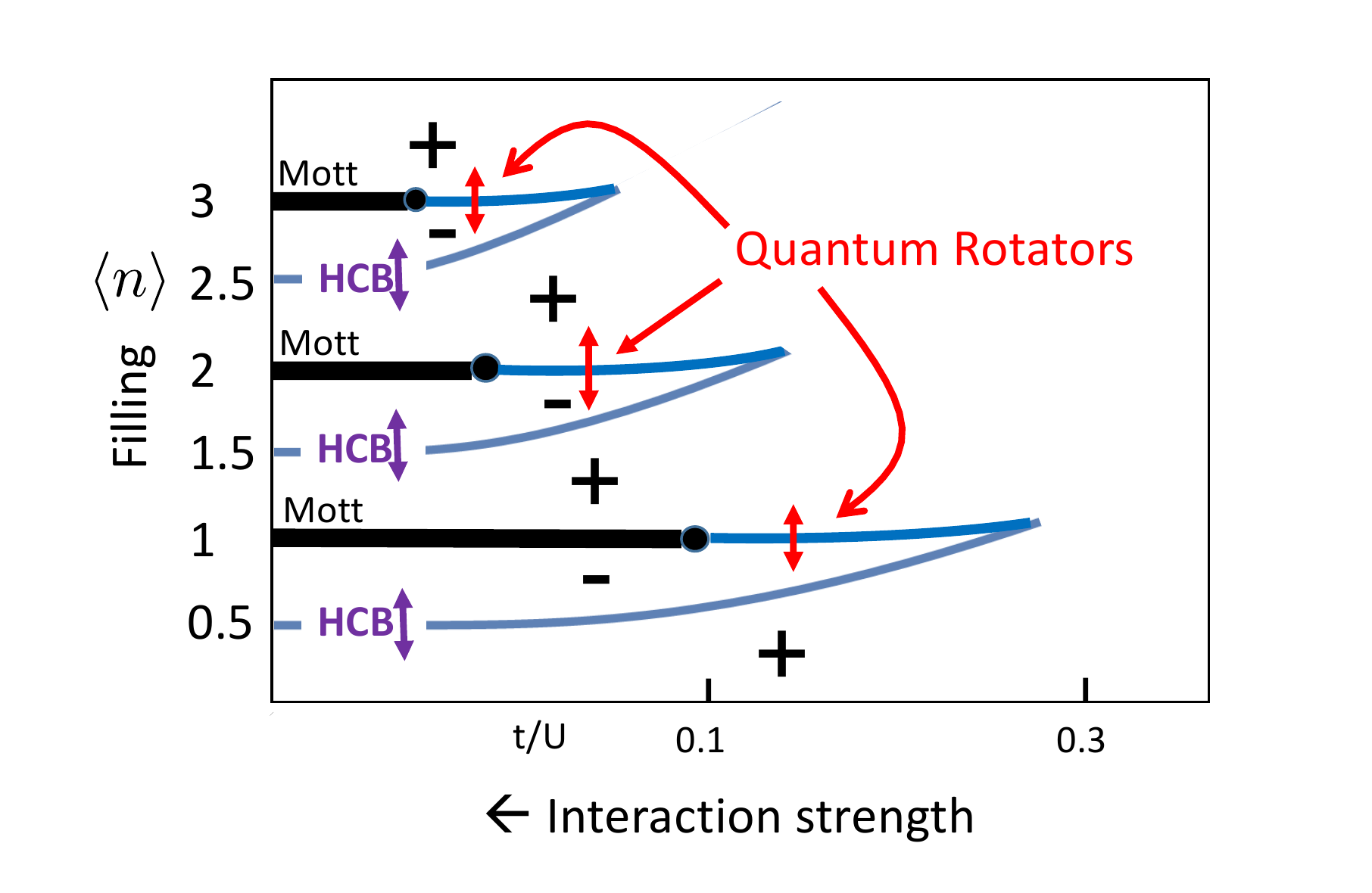}
\caption{Hall signs in strong interactions regime of the Bose Hubbard model Eq.~(\ref{BHM}).  Mott insulators are thick black lines, ending at critical points (black circles). 
Solid blue lines mark Hall sign changes at zero temperature, computed by Huber and Lindner\cite{huber}.
At high temperatures, we find the same sign changes using Quantum Rotators, and Hard Core Bosons (HCB) in Eqs.~(\ref{QR},\ref{Sign}).
}
\label{fig:PD}
\end{center}
\end{figure}

{\em Weak disorder and interactions.} ---  The f-sum rule (gauge invariance) yields
$\tau_{xx} 
\!=\! {n (e^*)^2 \over m}$.  Thus, using ( \ref{identities}) in (\ref{Formula}),  Drude's result is obtained at the zeroth order: $ n_H^{(0)} \!=\!n$.
Higher order terms in (\ref{Formula}) are suppressed by a common factor 
\be
\left( {\Delta_1\over \Delta_2}\right)^2  \propto   { \sum_\bq q^2_x |\Phi_\bq|^2 \kappa_\bq \over \epsilon_F }\ll 1,
\ee
where $\kappa_\bq$ is the wavevector dependent isothermal compressibility, and $\epsilon_F$ is the Fermi energy. 
Thus the sum in Eq.~(\ref{Formula}) produces systematic corrections to Drude theory due to potential fluctuations and interactions.

 {\em Strong interactions.} --- In the presence of a large Mott-Hubbard gap, induced by strong interactions, and at low temperatures, one can
 replace the operators $H$, $M$ and $\bj$, in Eq. (\ref{dsdB}), and thus in Eq.~(\ref{Formula}),  by renormalized effective Hamiltonian and  its derivatives~\cite{Com2}.  
 The Krylov states, recurrents and magnetization matrix elements are modified accordingly.  Formula (\ref{Formula}) can then by computed for the effective Hamiltonians, such as the Hubbard, t-J \cite{IEQM} and Kondo lattice models\cite{hewson}. These are relevant to strongly correlated metals, including the normal phase of unconventional superconductors, and Heavy Fermion phases.  
 The Hall number of these modes will be investigated elsewhere~\cite{Ilia}.

Here we study the Bose Hubbard model (BHM),  
\be
H^{\rm BHM}= -t\sum_{\langle ij \rangle} e^{-ie^*A_{ij}} a^\dagger_i a_j + \mbox{h.c.} + {U\over 2} \sum_i n_i^2 -\mu n_i ,
\label{BHM}
\ee
where $a^\dagger_i$ creates a lattice boson on site $i$, with occupations $n_i\!=\!0,1,\ldots$. The BHM is relevant to superconducting Josephson junction arrays, and to cold atoms in optical lattices.
At large  $U/t$, there are gapped Mott insulator phases at  integer fillings $\langle n_i\rangle =\mbox{integer}$. 
Huber and Lindner\cite{huber} have computed the ground state Chern number on finite tori. Here we
obtain the finite temperature Hall number sign for the thermodynamic metal, and compare it to the Chern calculations as shown in Fig.~\ref{fig:PD}. 

1. Near the superfluid to Mott insulator critical points at integer fillings $n_0$,  we replace $H^{\rm BHM}$ by Quantum Rotators (QR),
\begin{widetext}
 \be
H^{\rm QR} = \int d^d x  {1\over 2\chi_c} (\rho(\bx)-n_0 a^{-d})^2+ {1\over 2} \rho_s \left(\nabla\varphi(\bx) +{e^*\over c} \bA\right)^2 (1+ \gamma \rho(\bx)^2) +  \Phi(\bx)\rho(\bx).
\label{Hrotor}
\ee
\end{widetext}
$a$ is the lattice constant, $\chi_c$ is the local compressibilty, and $\rho_s$ is the local superfluid stiffness.  
$\gamma >0 $ since the superfluid order parameter increases away from the Mott phases. 
The canonical density-phase commutations are\cite{Com3} ,
\be
[\rho(\bx),\varphi(\bx')] = -i \delta(\bx-\bx').
\label{rhophi}
\ee
The  QR currents and magnetization densities are,
\bea
\bj (\bx) &=& - e^* \rho_s  \nabla\varphi  (1+ \gamma \rho^2),\nonumber\\
m(\bx)&=& - {e^*\over 2c} \left(  x j^y(\bx) -   y j^x(\bx)\right).
\eea
Thus we can evaluate the sign of the leading term as,
\bea
 {1\over n^{(0)}_H }  \propto      {\gamma  \over  e^* \rho_s c }  \langle ( \rho a^d-n_0 ) \rangle + \cO\langle(\rho a^d-n_0)^2\rangle .
 \label{QR}
 \eea
The Hall number near the Mott  critical point changes sign in the same direction as determined  at  zero temperature using Chern numbers, as shown in Fig.~\ref{fig:PD}. 
Higher order terms in (\ref{Formula}) are suppressed in disorder free systems.

2.  Near half odd integer fillings, between Mott phases,  we can use the effective Hard Core Bosons  (HCB) model~\cite{lindner},
\be
H^{\rm HCB} = -t \sum_{\langle ij\rangle } e^{-ie^*A_{ij} } S^+_i S^-_j + \mbox{h.c.}
\label{HCB}
\ee
where $\bf{S}$  are effective spin half operators. $S_i^+$ creates a HCB at site $i$, and $S_i^z\!=\!n_i-{1\over 2}$ measures its occupation relative to half filling.

The HCB  currents  and magnetization are,
\bea
j^\alpha &=&    -i  e^* t  \sum_i \left(  e^{-ie^*A_{ii+\alpha}} S^+_i S^-_{i+\alpha} -  \mbox{h.c.} \right) , \nonumber\\
M&=&   {e^*\over 2 } \sum_{i }  x_i j^y_{i+y} -  y_i j^x_{i,i+x}.
\eea
Expanding (\ref{product}) at high temperature yields,
 \be
 (A|B) = \beta \Tr \rho_\infty A^\dagger B -  {\beta^2\over 2} \Tr \rho_\infty \{H,A^\dagger\} B + \cO(\beta^3).
 \ee
The infinite temperature density matrix $\rho_\infty $ projects onto a fixed particle number  $\sum_i S^z_i=(n-{1\over 2}) V$.
$ \tau_{xx}  \!= \!  \beta  \Tr \rho_\infty  j_{i,i+x}^2 $. The traces in the magnetization matrix elements $M_{2j,2k}''$ vanish unless the operators encircle a magnetic flux. 
Therefore, for a {\em triangular} lattice at high temperatures,   $M_{0,0}'' \!\propto \! - \beta   (n-{1\over 2})$, while
for a square lattice   $M''_{0,0}\!\propto\! -\beta^2 (n-{1\over 2})$. 
Thus we obtain,
\be
{1\over n_H^{(0)} } \propto  \left\{ \begin{array}{ll}- T (n-{1\over 2}) & \mbox{triangular}\\  -  (n-{1\over 2}) & \mbox{square}\\ \end{array} \right.
\label{Sign}
\ee
High  order terms  include  $M_{2j,2k}''$, which 
 decay  rapidly with $j,k$ due to diminishing overlaps between Krylov states. Thus the Hall sign of HCB, in Eq.~(\ref{Sign}), is depicted in Fig.~\ref{fig:PD}.
We note that lattice effect resembles the behavior at infinite frequency\cite{Shastry}.

{\em Summary} -
Eq.~(\ref{Formula}) provides an exact computable formula for the Hall number of metals where $\infty\!>\!\sigma_{xx}\!>\!0 $. It should prove useful for numerical studies of 
disordered and strongly correlated, non relativistic fermions and bosons.  
The formula does not require well defined quasiparticles, as needed for Boltzmann's equation. It also circumvents 
numerical difficulties associated with real-time response functions, such as the Kubo formulae for conductivities. We look  forward to its application in experimentally relevant models of 
strongly correlated electron systems.

{\em Acknowledgements} I thank Yosi Avron, Ilia Khait, Netanel Lindner, and Ari Turner for useful discussions. 
I acknowledge support from the
US-Israel Binational Science Foundation grant 2016168 and  the Israel Science Foundation grant 2021367. I thank the Aspen Center for Physics,  grant NSF-PHY-1066293, and Kavli Institute for Theoretical Physics at Santa Barbara, where part of this work was done.
\bibliographystyle{unsrt}

\bibliography{HallN}

\begin{thebibliography}{10}
\bibitem{Comm-limits}
The correct order of limits is that the frequency scale of the current, and the
  $i0^+$ limit, are taken to zero after the thermodynamic limit ${N,V}\to
  \infty$.
\bibitem{Com-B}
The ambiguity in $\left({1\over \cL(B)}\right)' j^\alpha$, (due to arbitrary
  addition of operators which commute with {$H$}), is eliminated by matching
  the matrix elements at high energies ({$E_n,E_m\!\gg\! | \Phi | $} as:
  {$j^\alpha_{nm}/(E_n-E_m)\to {i \over \hbar \omega_c}\epsilon_{\alpha\beta}
  j_{nm}^\beta $}. {T}his asymptotic matching respects the semiclassical
  correspondence principle.
\bibitem{Com2}
Using renormalized operators is not advisable for the derivation of
  {E}q.~(\ref{Formula}), which is based on the {\em microscopic} identities of
  {E}q.~(\ref{identities}).
\bibitem{Com3}
The minimal coupling {$\nabla\varphi(\bx) +{q\over c} \bA$} is consistent with
  the sign of the density-phase commutator.
\end{thebibliography}

\newpage
\clearpage
\appendix
\centerline{\Large \bf Supplemental Material}
\section{Kubo Formulae}
The Kubo formula for the zero wavevector  dynamical conductivity   is  
\begin{widetext}
\be
\sigma_{\alpha\beta}(\omega) =\Im  \left( {1\over  V  Z  }\sum_{n,m}  \left( {e^{-\beta E_n} - e^{-\beta E_m} \over  (E_m-E_n)  (E_m-E_n -\hbar\omega-i0^+)} \right)  \langle n| j^\alpha  |m\rangle \langle m|j^\beta |n\rangle \right)   
\ee
\end{widetext}
$Z=\Tr e^{-\beta H}$, and $E_n, |n\rangle$ are the eigenenergies and eigenstates of $H$, respectively.

It is convenient to write the Bogoliubov inner product as a trace,
\bea
(A|B)&=& \Tr (\cW A^\dagger) B \nonumber\\
\cW \bullet  &\equiv &  - \left[\rho, \left({1\over \cL}\right)' \bullet \right]
\eea

It is easy to verify that in this hyperspace, $\cL ,\cW$ commute and therefore
\bea
(A|\cL|B) &=&  \Tr (\cW A^\dagger) \cL B = - \Tr (\cL \cW \cL B^\dagger) A \nonumber\\
&=&    \Tr ( \cW (\cL B)^\dagger) A= (\cL B|A),
 \eea
which proves that $\cL$ is hermitian in Bogoliubov hyperspace.
The Liouvillian resolvent includes an $i0^+$ 
prescription, and can be separated into  hermitian and antihermitian parts: $\left({1\over \cL -\omega - i0^+} \right)= \left({1\over \cL -\omega} \right)'+ i \left({1\over \cL -\omega - i0^+} \right)''$.

The longitudinal conductivity  is written as a matrix element in hyperspace:
\bea
\sigma_{xx}(\omega) &=&   { \hbar \over V} \Im  \left( j^x \Big|   \left( {1\over \cL-\hbar \omega-i0^+}\right) \Big|j^x\right)\nonumber\\
&=&  { \hbar \over V}    \left( j^x \Big|    \left( {1\over \cL-\hbar \omega-i0^+}\right)'' \Big|j^x\right)
\label{sxx-kubo}
\eea
The Hall conductivity is given by taking the imaginary part of $\langle n| j^\alpha  |m\rangle \langle n|j^\beta |m\rangle$
 and the real part of $1/ (E_n-E_m-\hbar \omega - i0^+ ) $.
  \be
 \sigma_{H}(\omega) ={ \hbar \over V} \Im  \left( j^x \Big|  \left( {1\over \cL-\hbar \omega }\right)' \Big|j^y\right)
 \label{sigmaH}
 \ee 
It is easy to verify that $\sigma_H(\omega)$ is antisymmetric in $x\leftrightarrow y$,  because of hermiticity of $j^x,j^y$.  

Note: the symmetric (dissipative) part is produced by the $\delta$-function contributions,
 \bea
 \sigma_{xy}^{\rm symm} ( \omega)&=& { \hbar \over V} \Re  \left( j^x \Big|   \left( {1\over \cL-\hbar \omega-i0^+ }\right)'' \Big|j^y\right)\nonumber\\
 &=& \pi   {\hbar \over  V  Z  }\sum_{n,m} {e^{-\beta E_n} -e^{-\beta E_m}\over E_m-E_n }\nonumber\\
 && \times \Re\left( \langle n| j^x |m\rangle \langle m|  j^y  |n\rangle \right) \delta(E_m-E_n-\hbar\omega)\nonumber\\
&=&   \sigma_{yx}^{\rm symm} (\omega)
 \label{sigma-symm}
 \eea
We can set $ \sigma_{xy}^{\rm symm}=0$, and $\sigma_{xx}=\sigma_{yy}$, assuming C4 symmetry of the (disorder averaged) Hamiltonian.
For $\omega \to 0$, Eq.~(\ref{sxx-kubo}) and (\ref{sigmaH}) reduce to the conductivities equations (4) in the main text.

\section{Krylov states and Recurrents}
\label{App:CF}

For the current-current response functions,  an orthonormal Krylov basis is constructed  from root hyperstates by Eq.~(16).
The {\em recurrents} are the `hopping'' matrix elements between Krylov states, depicted in Fig. 1 of the main text.
\be
\Delta_n =  \langle n+1,\alpha |\cL |n,\alpha\rangle = L_{n+1,n}
\ee

The conductivity moments are defined by,
\bea
\mu_{2k} &=&  \int_0^\infty {d\omega\over \pi} \omega^{2k} \sigma_{xx}(\omega)\nonumber\\
&=&   \tau_{xx}  \left({L\over \hbar}\right)^{2k}_{0,0}\nonumber\\
&=&   \tau_{xx} \left( \begin{array}{cccc}
  0&     \Delta_1& 0 & \ldots\\
   \Delta_1& 0  &    \Delta_2&  \ldots\\
0&    \Delta_2&  0 &   \\
\vdots& \vdots &   &\ddots
\end{array}
\right)^{2k}_{0,0} 
\eea
which yields recursion relations between the $n$ lowest (even) moments $\mu_{2k},k=1,\ldots n$,  and the $n$ lowest recurrents $\Delta_k, k=1,n$, e.g.
\bea
\mu_0&=&  \tau_{xx}\nonumber\\
{\mu_2\over \mu_0}&=& \Delta_1^2\nonumber\\
{\mu_4\over \mu_0}&=&  \Delta_1^2(\Delta_1^2+\Delta_2^2)\nonumber\\
{\mu_6\over \mu_0} &=&\Delta_1^2\left( \Delta_1^4 + 2\Delta_1^2\Delta_2^2+\Delta_2^4+\Delta_2^2\Delta_3^3\right)\nonumber\\
\vdots&=&\vdots
\eea

We note that $j^\alpha$ is hermitian, and  $\cL$ transforms hermitian to antihermitian operators, and vice versa. Hence, 
the Krylov operators $|2j,\alpha\rangle$ ($|2j+1,\alpha\rangle$) are hermitian (antihermitian).
As $\Delta_n$ are given by traces of products of  two hermitian (antihermitian) operators, they are real numbers.
The complex Liouvillian Green function is the inverse of a tridiagonal matrix,
\be
G_{n, m}(z) =  \left( \begin{array}{cccc}
  z&   - \Delta_1& 0 & \ldots\\
-  \Delta_1& z  & -  \Delta_2&  \ldots\\
0&  - \Delta_2&  z &   \\
\vdots& \vdots &   &\ddots
\end{array}
\right)^{-1}_{n,m} 
\ee

 \subsection{Off diagonal response functions}
We can write the off-diagonal matrix elements of $G_{0,n}$, in terms of  $G_{0,0}$ and $\{\Delta_n\}$.
Setting $z=0$, from $-G L =1$, we obtain
\be
 - G_{0,1} \Delta_1 = 1,
 \ee
 and 
 \be
G_{0,n+2} =  - G_{0,n} { \Delta_{n+1} \over    \Delta_{n+2 } },
\ee
which yields recursion relations between terms of the same parity. Hence 
\bea
G_{0,2j}  &=& (-1)^{j} {\Delta_1 \Delta_3 \cdots \Delta_{2j-1} \over \Delta_2 \Delta_4 \cdots \Delta_{2j} }  G_{0,0}   , \nonumber\\
G_{0,2j+1}  &=&    (-1)^{j+1}   {\Delta_2 \Delta_4\cdots \Delta_{2j}\over \Delta_1 \Delta_3 \cdots ~\Delta_{2j+1} }  ,
\label{Gn0}
\eea
$G_{0,2j+1}(\omega)$ are purely real in the limit of $\omega\to 0$, and cannot contribute to the  imaginary parts of the matrix elements of
$\left({1\over \cL}\right)''$ in Eq.~23, of the main text.

The Kubo formula for $\sigma_{xx}$ can be used to verify that,
\bea
\Re \sigma_{xx}(\omega) &=&\Re \sigma_{xx}(-\omega) \ge 0 \nonumber\\
\Im \sigma_{xx}(\omega) &=&  -  \Im \sigma_{xx}(-\omega)  
 \eea
Hence,  $G_{0,0}(\omega\!=\!0)=i G_{0,0}''$ is purely imaginary.    
This implies that the non zero contributions to  Eq.~23, of the main text, are the even terms which contain $G''_{0,0}$ which, as seen in Eq.~21 of the main text,  
is proportional to the DC conductivity $\sigma_{xx}$.

 \end{document}